\documentclass[aps,prb,showpacs,twocolumn,amssymb,floatfix]{revtex4}

\usepackage{amsmath}
\usepackage{graphicx}
\usepackage{longtable}

\begin{document}
  \title{Electron dynamics in vacancy islands}
  \author{H.\ Jensen}
  \author{J.\ Kr\"oger}\email{kroeger@physik.uni-kiel.de}
  \author{R. Berndt}
  \affiliation{Institut f\"ur Experimentelle und Angewandte Physik, Christian-Albrechts-Universit\"at zu Kiel, D-24098 Kiel, Germany}
  \author{S.\ Crampin}\email{s.crampin@bath.ac.uk}
  \affiliation{Department of Physics, University of Bath, Bath BA2 7AY, United Kingdom}

  \begin{abstract}  The dynamics of Ag(111) surface state electrons confined to
    nanoscale hexagonal and triangular vacancy islands are investigated using
    scanning tunneling spectroscopy. The lifetimes of quantised states with
    significant amplitude near the centers of the vacancies are weakly affected by
    the geometry of the confining cavity.  A model that includes the dependence of
    the lifetime on electron energy, vacancy size, step reflectivity and the phase
    coherence length describes the results well.  For vacancy islands with areas
    in the range $\approx 40$--$220\,{\rm nm}^2$ lossy scattering is the dominant
    lifetime-limiting process.
    This result and a corrected analysis of published experimental data improve the
    consistency of experimental and calculated surface state lifetimes.
  \end{abstract}


  \maketitle

  Spectacular real-space observations of electron confinement have been realized by
  scanning tunneling microscopy (STM) and scanning tunneling spectroscopy (STS) for
  the Shockley-type electronic surface states on the (111) surfaces of the noble
  metals. Electron confinement to artificial ({\it e.g.}, Refs.\
  \onlinecite{cro93b,hel94_,kli00a,kli01_,bra02_}) and natural ({\it e.g.}, Refs.\
  \onlinecite{jli98b,jli99_,bur98_,bar03_}) nanostructures has been studied in some
  detail. Moreover, several reports have addressed the lifetime of surface state
  electrons as deduced from tunnelling spectroscopy of large terraces, scattering
  patterns near steps and in confining structures.
  \cite{jli98b,bra02_,bur99_,kli00b,vit03_} Currently, however, the influence of
  the confinement on electronic lifetimes is not clear.

  Using low-temperature STS of triangular and hexagonal vacancy islands and model
  calculations, we investigate the effects of the geometry of the confining
  resonator and of lossy scattering at its boundary on the lifetimes of confined
  states.  For the states investigated, which have an antinode at the center of
  these resonators, we find similar lifetimes independent of the geometry. Lossy
  scattering at the confining step edges turns out to be the lifetime-limiting
  process.

\begin{figure}
    \includegraphics[bbllx=10,bblly=440,bburx=575,bbury=730,width=60mm,clip=]{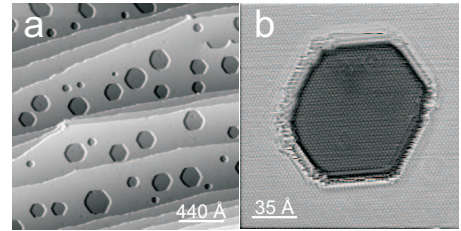}
    \caption[topo]{(a) Topographic image of Ag(111) surface after low-flux Ar$^{+}$
    bombardment. (b) Atomically resolved vacancy island which serves as the cavity
    for electron confinement.} \label{topo}
  \end{figure}

  The experiments were performed in ultrahigh vacuum (base pressure $1\times
  10^{-8}\,{\rm Pa}$) at $9\,{\rm K}$ using a custom-made microscope. Tungsten tips
  and Ag(111) were prepared by argon ion bombardment and annealing. Hexagonal
  vacancy islands were fabricated by briefly ($2-3\,{\rm s}$) exposing  freshly
  prepared Ag(111) surfaces to a low-flux argon ion beam. \cite{mor95_} The
  resulting nanostructures are shown in Fig.\ \ref{topo}. Triangular vacancy
  islands were obtained through controlled tip-sample contacts.
  The differential conductivity (${\rm d}I/{\rm d}V$) was measured by superimposing
  a sinusoidal voltage modulation ($3\,{\rm mV}_{\rm rms}$, $10\,{\rm kHz}$) on the
  tunneling voltage and measuring the current response by a lock-in amplifier.

\begin{figure}
    \includegraphics[bbllx=20,bblly=90,bburx=585,bbury=775,width=60mm,clip=]{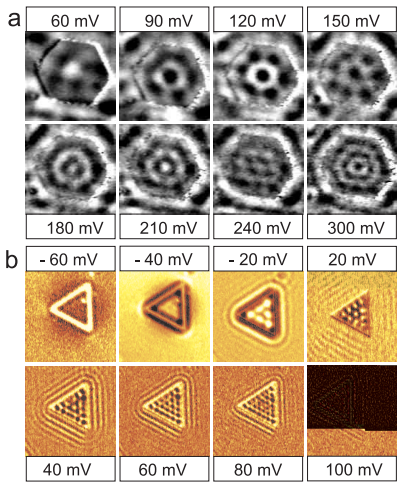}
    \caption[didv]{(a) Series of constant-current ${\rm d}I/{\rm d}V$ maps of a hexagonal vacancy island with edge length
    $\approx 5.8\,{\rm nm}$ recorded at indicated sample voltages (white corresponds to large ${\rm d}I/{\rm d}V$).
    (b) Analogous data for a triangular vacancy with edge length $22\,{\rm nm}$.}
    \label{didv}
  \end{figure}

  Maps of ${\rm d}I/{\rm d}V$ from a hexagonal vacancy island (Fig.\ \ref{didv}a)
  exhibit strongly voltage-dependent features, similar to those found from, {\it
  e.g.}  hexagonal Ag islands. \cite{jli98b,jli99_} For voltages below the Ag(111)
  surface state band edge at $E_0\approx -67\,{\rm\ mV}$ the interior of the
  vacancy island is featureless while at higher voltages increasing numbers of
  sixfold azimuthally symmetric rings occur. The sixfold modulation is partially
  distorted due to deviations of the vacancy island from a perfect hexagonal shape.
  ${\rm d}I/{\rm d}V$ being related to the local density of states,\cite{jli97_}
  the interference patterns are attributed to surface state standing waves confined
  by scattering at the edges of the island. Figure \ref{didv}b shows similar data
  for a triangular vacancy island. Starting from a featureless interference pattern
  at $-60\,{\rm mV}$, increasingly complex patterns occur at higher voltages. The
  number of antinodes increases with higher voltages as expected from a
  particle-in-a-box model. At $-20\,{\rm mV}$, for instance, six antinodes of the
  local density of states are clearly visible.

\begin{figure}
    \includegraphics[bbllx=80,bblly=105,bburx=500,bbury=570,width=45mm,clip=]{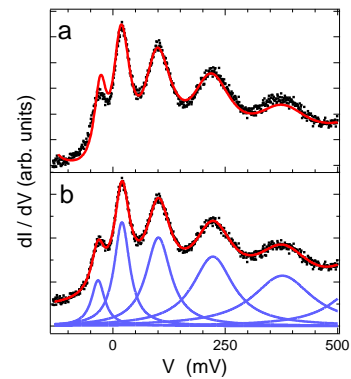}
    \caption[spec1]{(a) Experimental ${\rm d}I/{\rm d}V$ data (dots) fitted by calculated spectrum (full line) based
    on a variation calculation as explained in the text. (b) Experimental ${\rm d}I/{\rm d}V$ spectrum (dots) fitted by
    superposition of Lorentzians (full lines). Topographic edge length of the island is $7.4\,{\rm nm}$.}
    \label{spec1}
  \end{figure}

  For a quantitative measure of the confinement we acquired ${\rm d}I/{\rm d}V$
  spectra above the vacancy island centers. A typical spectrum of a hexagonal
  island (Fig.\ \ref{spec1}a) is comprised of a series of peaks with the first
  appearing just above the lower band edge of the Ag(111) surface state. The peak
  width increases with increasing energy. These peaks correspond to
  the energy levels of the confined Ag(111) surface state electrons, with the
  broadening into resonances as a result of single-particle scattering
  processes,\cite{hel94_,cra94a} many-body interactions,\cite{cra96_} and
  instrumental effects such as the finite modulation voltage and finite
  temperature. We have confirmed this interpretation using the same approach as in
  Refs.\ \onlinecite{jli98b,jli99_} using a variational embedding technique
  \cite{cra95_} to compute the electronic structure of two-dimensional electrons
  with effective mass $m^{\ast}$ confined to hexagonal or triangular domains by an
  infinite barrier. This gives a series of discrete levels at energies that are
  dependent upon the size and geometry of the confined region, which we broaden by
  including an empirically determined self-energy ${\text{Im}}\,\Sigma\approx
  0.2\,(E-E_{\text{F}})$, where $E_{\text{F}}$ is the Fermi energy. Note that in
  comparing the experimental data with calculated spectra the true dimensions of
  the island must be determined. It is not {\it a priori} clear that the relevant
  dimension of the island coincides with its apparent topographic boundaries, for
  example due to the finite spatial extension of the tip the topographic dimensions
  are expected to be smaller than the actual ones. Consequently, we fit the
  experimental ${\rm d}I/{\rm d}V$ curves by calculated spectra varying the edge
  length as the only fit parameter. The full line in Fig.\ \ref{spec1}a depicts the
  result of this fit procedure applied to the vacancy island with a topographic
  edge length of $7.4\,{\rm nm}$. The fit parameter, however, is given by
  $7.6\,{\rm nm}$. Performing this fit procedure for several island sizes we find
  that the topographs of the vacancy islands tend to underestimate the actual
  sizes.

\begin{figure}
    \includegraphics[bbllx=10,bblly=0,bburx=420,bbury=410,width=65mm,clip=]{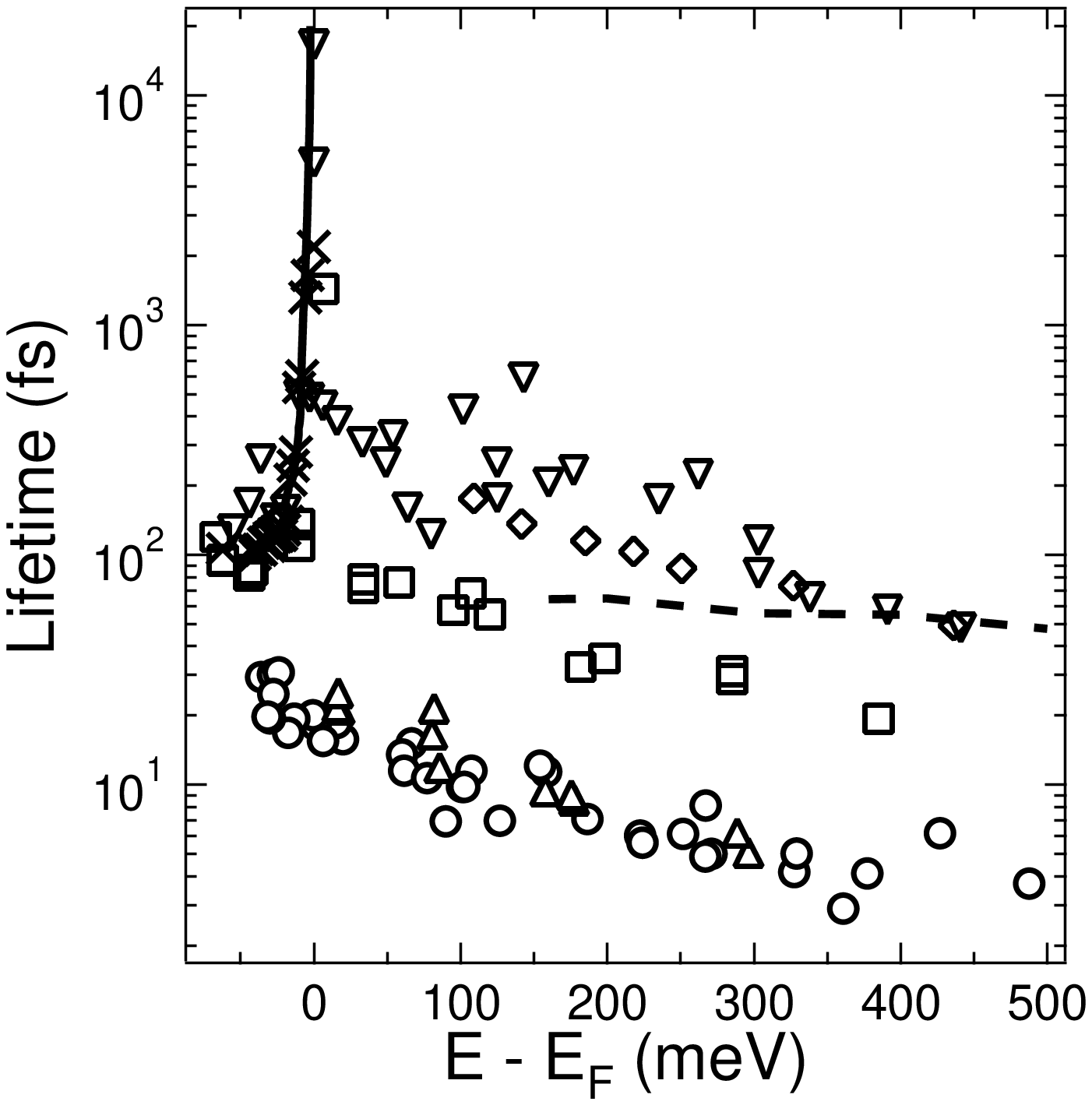}
    \caption[tau]{Lifetime versus binding energy $E - E_{\rm F}$ found for Ag(111) surface state electrons confined
    to vacancy hexagons ($\circ$), vacancy triangles ($\vartriangle$), triangular corrals of Ag adatoms ($\triangledown$,
    adapted from Ref.\ \onlinecite{bra02_}), circular and rectangular corrals of Mn adatoms ($\square$, adapted from
    Ref.\ \onlinecite{kli01_}); lifetimes extracted from standing wave patterns near step edges ($\lozenge$, adapted from
    Refs.\ \onlinecite{vit03_,bur99_}); photoemission data ($\times$, from Ref.\ \onlinecite{eig02_}); results of theory
    are depicted as a full and a dashed line (from Refs.\ \onlinecite{eig02_} and \onlinecite{vit03_}, respectively).}
    \label{tau}
  \end{figure}

  We now discuss the lifetimes of the quantised electron states which are proportional to the inverse
  width of the associated peaks in the ${\rm d}I/{\rm d}V$ spectra. In particular the half width at half maximum  (HWHM)
  of the electron levels corresponds to an energy uncertainty (${\text{Im}}\,\Sigma$)
  related to the lifetime $\tau$ of the electronic state via
  \begin{equation}
    \tau = {\rm\hbar} / (2\,{\text{Im}}\Sigma) \approx \left[\,329\,{\rm meV\,fs}\,\right] / {\text{Im}}\,\Sigma.
    \label{eqn:tausig}
  \end{equation}
  In analysing the peak width, instrumental broadening due to voltage modulation
  and thermal effects have been taken into account. A fit using 5 Lorentzians plus
  an additional one at high energies to match the background is shown in Fig.\
  \ref{spec1}b. Taking the HWHM of the Lorentzians we arrive at the corresponding
  imaginary parts of the self energy, which is converted to a lifetime using
  (\ref{eqn:tausig}). Figure \ref{tau} displays the electronic lifetimes versus
  their binding energies as extracted from various hexagonal and triangular vacancy
  islands, along with data from several other Ag(111) surface state lifetime
  studies, both experimental and theoretical. While a general trend of increasing
  lifetime as the binding energy approaches the Fermi level is obvious, there are
  significant differences in magnitudes, to be discussed next.

  We first note from the vacancy data of Fig.\ \ref{tau} that the geometric shape of the confining structure
  has little impact on lifetimes. Consequently, in modeling, we consider a {\emph{circular}} vacancy island
  where the high symmetry permits a relatively simple description. In particular,
  one can show \cite{cra04b} that the confined levels occur at energies for which
  \begin{equation}
    R\,{\rm e}^{2{\rm i}kS}\,{\rm e}^{-S/L_{\Phi}}\,{\rm e}^{-{\rm i}\pi/2} = 1.
    \label{eqn:roundtrip}
  \end{equation}
  $S$ is the radius of the vacancy, $k=\sqrt{2m^{\ast}(E-E_0)/\hbar^2}$ is the magnitude of the surface electron
  wave vector, and $R$ is the reflection coefficient describing scattering from the confining atomic step.
  Eq.\ \ref{eqn:roundtrip} is similar to the round trip phase condition for surface
  states in the phase accumulation model, \cite{ech78_,smi85_} where electrons
  undergoing multiple reflections at barriers with reflection coefficients
  $R_{\text{l}}$ and $R_{\text{r}}$ separated by distance $d$ are found at energies
  given by \(R_{\text{l}}\,R_{\text{r}}\,\exp\,2{\rm i}kd=1\). In the vacancy
  island the round trip corresponds to starting and finishing at the center and
  only involves one reflection --- so just one reflection coefficient appears in
  (\ref{eqn:roundtrip}) --- and in the vacancy island the wave function must have
  an antinode at the center to be visible in STS, so that the round trip phase
  corresponds to an odd number of half wavelengths. This is the effect of the
  $\exp\,{\rm -i}\pi/2$ factor in (\ref{eqn:roundtrip}). One final difference is
  that (\ref{eqn:roundtrip}) also takes into account many-body interactions
  through the factor containing $L_{\Phi}$, the phase relaxation length due to
  electron-electron (e-e) and electron-phonon (e-p) scattering. $L_\Phi$ is related
  by the group velocity $v_{\text{g}}=\hbar k/m^{\ast}$ to the corresponding
  inelastic scattering lifetime $\tau_{\text{I}}$: $L_\Phi=v_{\text{g}}
  \tau_{\text{I}}$.

  Writing $R=\exp{\rm i}\left(\phi_R - {\rm i}\ln |R|\right)$, Eq.\ \ref{eqn:roundtrip} is satisfied when
  \begin{equation}
    \phi_R - {\rm i}\ln |R| + 2\,k\,S + {\rm i}S/L_{\Phi} - \pi/2 = 2\pi\,n
    \label{eqn:complexphase}
  \end{equation}
  for $n=1,2,\ldots$,
  which has solutions at complex energies $E - {\rm i}\,{\text{Im}}\,\Sigma$. Assuming ${\text{Im}}\,\Sigma\ll (E-E_0)$,
  the real and imaginary parts of Eq.\ \ref{eqn:complexphase} give the energy of the  $n$'th level
  \begin{equation}
    E_n = E_0 + \frac{{\rm\hbar}^2(n\pi + \pi/4 - \phi_R/2)^2}{2m^{\ast}\,S^2}
  \end{equation}
  and corresponding width
  \begin{equation}
    {\text{Im}}\,\Sigma_n = \frac{{\rm\hbar}\,v_{\text{g}}}{2}\left[ -\frac{\ln |R|}{S} + \frac{1}{L_{\Phi}}\right]
    \label{eqn:sig}
  \end{equation}
  where $v_{\text{g}}$, $|R|$ and $L_{\Phi}$ are evaluated at the energy $E_n$.
  Hence the lifetime of the surface state electrons confined within the vacancy islands
  is given by
  \begin{equation}
    \tau^{-1} = \tau_R^{-1} + \tau_{\text{I}}^{-1}
    \label{eqn:tau}
  \end{equation}
  where the lifetime $\tau_R$ associated with lossy scattering at the atomic step is
  \begin{equation}
    \tau_R = -\frac{S}{v_{\text{g}}\ln |R|}.
    \label{eqn:taur}
  \end{equation}
  At $225\,{\rm meV}$, Echenique and co-workers \cite{vit03_} find
  $\tau_{\text{I}}\approx 62\,{\rm fs}$ in calculations that include the surface
  band structure, and which treat e-e interactions within the GW approximation
  \cite{kli00b} and use the full Eliashberg spectral function for the e-p
  interaction. \cite{eig02_} B\"urgi {\it et al.} \cite{bur98_} have measured the
  reflection coefficient for surface state scattering at ascending steps on Ag(111)
  and report $|R|\approx 0.23\pm 0.07$ at $225\,{\rm meV}$.
  \begin{figure}
    \includegraphics[bbllx=100,bblly=90,bburx=460,bbury=350,width=45mm,clip=]{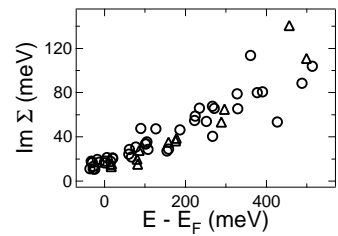}
    \caption[sigma]{Hexagon ($\circ$) and triangle ($\vartriangle$) data from Fig.\ \ref{tau}, represented
    as ${\rm Im}\,\Sigma$ versus binding energy $E-E_{\rm F}$.}
    \label{sigma}
  \end{figure}
  Using Eq.\ \ref{eqn:taur} (and $m^{\ast}=0.4\,m_{\rm e}$)\cite{bur99_} this
  gives $\tau_R\approx (9\pm 2)\,{\rm fs}$ for an
  island with edge $7.5\,{\rm nm}$, and from (\ref{eqn:tau}) a lifetime of
  $\tau\approx (7.9\pm 1.5)\,{\rm fs}$. This is consistent with our measured
  lifetimes shown in Fig.\ \ref{tau}. The calculated result is exemplary for the
  overall consistency with our measured lifetime data. We can therefore conclude
  that at this energy the electron states within the vacancy island are roughly $8$
  times more likely to decay via lossy scattering at the confining boundary than
  via inelastic e-e and e-p scattering. Repeating this analysis at other energies,
  we find that the lossy boundary scattering is the dominant lifetime-limiting
  process for all islands that we have studied. They would need to be typically an
  order of magnitude larger in dimension for the inelastic e-e and e-p scattering
  to become dominant.

  B\"urgi {\it et al.} \cite{bur98_} find that $|R|$ decreases with increasing
  energy. Using their values we find that $-\ln |R|$ varies approximately as
  $(E-E_0)^{1/2}$. Since $v_{\text{g}}\propto (E-E_0)^{1/2}$ and since
  $L_{\Phi}^{-1}$ is negligible with respect to $\ln|R|/S$ in Eq.\ \ref{eqn:sig}
  due to the dominance of the lossy-scattering mechanism the level width increases
  approximately linearly with energy. Figure \ref{sigma} shows our results for the
  linewidths as a function of energy, which demonstrate this behavior. A similar
  variation in level width was previously inferred by Li {\it et al.} \cite{jli98b}
  for surface state confinement to adatom islands, where a phenomenological self
  energy ${\text{Im}}\,\Sigma=0.2\,(E-E_0)$ was found to match the observed level
  broadening. B\"urgi {\it et al.}\cite{bur98_} have shown that the reflection
  coefficient for descending steps, which surround adatom islands, varies in a
  similar manner to that for ascending steps.

  In Fig.\ \ref{tau} various other lifetime data are shown.  The Mn corral values
  obtained by Kliewer {\it et al.} \cite{kli01_} ($\square$) have been obtained by
  analysing experimentally determined spectral linewidths using the two-dimensional
  ``black-dot'' scattering model introduced by Heller and coworkers. \cite{hel94_}
  The lifetime is obtained using Eq.\ \ref{eqn:tausig} where $\text{Im}\,\Sigma$ is
  the electron self energy needed to bring calculated spectra into agreement with
  measured ones. In effect, the lifetime is assumed to be due to e-e/e-p
  scattering, and hence to be identified as $\tau_{\text{I}}$, as the scattering
  properties of the confining adatom array is incorporated via the
  multiple-scattering equations. However, the nature of the system is similar to
  the present one, with the total linewidth a sum of contributions due to lossy
  scattering and the e-e/e-p scattering.  Hence the assumption that Mn adatoms act
  as ``black-dot'' scatterers means that the lifetime values that have been
  obtained should be viewed as lower limits to $\tau_{\text{I}}$. The true
  reflectivity of the Mn adatom corrals is likely to be lower than that of the
  assumed ``black-dot'' scatterers, meaning that the lossy scattering is actually
  greater than modelled. Hence a smaller amount of the measured linewidth should be
  attributed to e-e/e-p scattering, increasing $\tau_{\text{I}}$.

  The lifetimes deduced from the decay of standing wave patterns
  \cite{bur99_,vit03_} at steps shown in Fig.\ \ref{tau} ($\lozenge$) are not
  complicated by lossy scattering as the asymptotic decay of the standing waves
  depends only upon $\tau_{\text{I}}$. However, B\"urgi {\it et al.} \cite{bur99_}
  have misidentified the correct dependence of the standing wave patterns on the
  phase coherence length, so that their are actually twice $\tau_{\text{I}}$.
  \cite{cra04a} The lifetimes determined by Braun and Rieder \cite{bra02_} shown in
  Fig.\ \ref{tau} ($\triangledown$) have been obtained from a detailed analysis of
  standing wave patterns in triangular corrals constructed from Ag adatoms. In
  their analysis the Ag adatoms are not treated as ``black-dot'' scatterers, but as
  point scatterers whose scattering properties are determined by fitting, along
  with a phase coherence length which enters an attenuation factor
  $\exp{-r/\tilde{L}_{\Phi}}$ in the electron propagation between scattering
  events, and to and from the STM tip position. Thus lossy scattering effects are
  fully accounted for in this work -- in so far as the two-dimensional
  point-scatterer model is valid -- but again an incorrect relationship is used
  between the phase coherence length $\tilde{L}_{\Phi}$ and the lifetime
  $\tau_{{\rm I}}$, leading to values that are twice $\tau_{\text{I}}$. Correcting
  for this, the agreement between the lifetimes found by Braun and Rieder and
  theoretical values for Ag(111), which are generally lower than those found in
  Ref.\ \onlinecite{bra02_}, is improved.

  In summary we have investigated the lifetimes of electrons confined to vacancy
  islands with areas  $\approx 40 - 220\,{\rm nm}^2$ on Ag(111). We find that the
  geometry of the vacancy has only a weak influence on the lifetimes, which are
  dominated by lossy scattering at the steps which form the edges of the island.
  The lifetimes are well described by a theory that includes the dependence on the
  electron energy, vacancy size, step reflectivity and the phase coherence length.
  This indicates that the crossover to lifetimes dominated by electron-electron and
  electron-phonon scattering will occur for islands with dimensions an order of
  magnitude greater than those studied here.  This result and a corrected analysis
  of published experimental data lead to a more consistent picture of surface
  state lifetimes in experiments and calculation.

  H. J., J. K., and R. B. thank the Deutsche Forschungsgemeinschaft for financial
  support. S. C. acknowledges the support of the British Council.

\end{document}